\documentclass[conference]{IEEEtran}
\IEEEoverridecommandlockouts

\usepackage{cite}
\usepackage{amsmath,amssymb,amsfonts}
\usepackage{algorithmic}
\usepackage{graphicx}
\usepackage{textcomp}
\usepackage{xcolor}

\usepackage{multirow}

\usepackage{xurl}
\usepackage{hyperref}

\def\BibTeX{{\rm B\kern-.05em{\sc i\kern-.025em b}\kern-.08em
		T\kern-.1667em\lower.7ex\hbox{E}\kern-.125emX}}
\begin{document}

		\begin{titlepage}
		\begin{center}	
			
			\huge
			\textbf{Conditional Generative Adversarial Networks Based Inertial Signal Translation}
			
			\vspace{0.5cm}
			\LARGE
			Accepted version
			
			\vspace{1.5cm}
			
			\text{Marcin Kolakowski }
			
			\vspace{.5cm}
			\Large
			 Institute of Radioelectronics and Multimedia Technology,\\Warsaw University of Technology,
			Warsaw, Poland,\\
			contact: marcin.kolakowski@pw.edu.pl

			\vspace{1.4cm}

		\end{center}
		
		\Large
		\noindent
		\textbf{Originally presented at:}
		
		\noindent
		2025 Signal Processing Symposium (SPSympo), Warsaw, Poland
		
		\vspace{.5cm}
		\noindent
		\textbf{Please cite this manuscript as:}

		\noindent
		M. Kolakowski, "Conditional Generative Adversarial Networks Based Inertial Signal Translation," 2025 Signal Processing Symposium (SPSympo), Warsaw, Poland, 2025, pp. 74-77, doi: 10.23919/SPSympo63739.2025.11124016.
		
		\vspace{.5cm}
		\noindent
		\textbf{Full version available at:}
		
		\noindent
		\url{https://doi.org/10.23919/SPSympo63739.2025.11124016}

		\vspace{.5cm}
		\noindent
		\textbf{Additional information:}
		
		\noindent
		Continuation of the study presented in:
		
		\vspace{0.2cm}
		
		\noindent
		M. Kolakowski, V. Djaja-Josko, J. Kolakowski, and J. Cichocki, “Wrist-to-Tibia/Shoe Inertial Measurement Results Translation Using Neural Networks,” Sensors, vol. 24, no. 1, p. 293, Jan. 2024, doi:  \url{https://doi.org/10.3390/s24010293} 
		
		\vspace{0.5cm}
		\noindent
		Associated data available at:
		
		\vspace{0.2cm}
		\noindent
		M. Kolakowski, “Wrist and Tibia/Shoe Mounted IMU Measurement Results for Gait Analysis.” Zenodo, Dec. 27, 2023. doi: \url{https://doi.org/10.5281/ZENODO.10436579}.

		\vfill
		
		\large
		\noindent
		© 2025 IEEE.  Personal use of this material is permitted.  Permission from IEEE must be obtained for all other uses, in any current or future media, including reprinting/republishing this material for advertising or promotional purposes, creating new collective works, for resale or redistribution to servers or lists, or reuse of any copyrighted component of this work in other works.
	\end{titlepage}

	\title{Conditional Generative Adversarial Networks Based Inertial Signal Translation\\
		\thanks{This research was funded by the Polish National Centre for Research and Development, grant number THCS/I/24/RENEW/2025.
		
		I would like to thank the Foundation for the Development of Radiocommunication and Multimedia Technologies for financial support in attending the conference.
				The author would like to thank the IoT Systems Research Group at Warsaw University of Technology for providing the sensors used for data collection. 
		}
	}
	
	\author{
		\IEEEauthorblockN{Marcin Kolakowski}
		\IEEEauthorblockA{\textit{Institute of Radioelectronics and Multimedia Technology} \\
			\textit{Warsaw University of Technology}\\
			Warsaw, Poland \\
			marcin.kolakowski@pw.edu.pl}
	}
	
	\maketitle
	
	\begin{abstract}
The paper presents an approach in which inertial signals measured with a wrist-worn sensor (e.g., a smartwatch) are translated into those that would be recorded using a shoe-mounted sensor, enabling the use of state-of-the-art gait analysis methods. In the study, the signals are translated using Conditional Generative Adversarial Networks (GANs). Two different GAN versions are used for experimental verification: traditional ones trained using binary cross-entropy loss and Wasserstein GANs (WGANs). For the generator, two architectures, a convolutional autoencoder, and a convolutional U-Net, are tested. The experiment results have shown that the proposed approach allows for an accurate translation, enabling the use of wrist sensor inertial signals for efficient, every-day gait analysis.
	\end{abstract}
	
	\begin{IEEEkeywords}
		signal translation, GAN, machine learning, inertial measurements, gait analysis
	\end{IEEEkeywords}

	\section{Introduction}
	In clinical settings, gait analysis is performed using laboratory-based motion capture systems. Unfortunately, despite their high accuracy, they are unsuitable for everyday use due to their high cost and stationary nature. A possible solution for everyday gait monitoring is using sensors equipped with inertial measurement units (IMUs). Those sensors can be placed in several areas of the human body, with lower back, feet, and shins being the most popular among the studies \cite{priscoValidityWearableInertial2025}. Unfortunately, wearing the sensors in those places may be uncomfortable and impractical. For example, shoe sensors may require sensor frequent reattachment when changing shoes or battery charging.

A more practical solution would be to use popular wrist wearables, e.g., smartwatches or smart bands. However, the validation studies reported that the methods used for wrist-based gait analysis are less accurate than those employed for sensors located in the lower body area \cite{priscoValidityWearableInertial2025}. The solution to that problem might be to translate the wrist-worn IMU measurement results to those that would be collected if the sensor was worn in the lower body area, e.g., strapped to a shoe. The obtained signals might then be processed with well-established lower body-area methods.

Translating the signals between the upper and lower limbs is a challenging task. Even simplified movement models take into account detailed, not easily measurable data such as limb segments' mass and dimensions \cite{olinskiDevelopmentSimplifiedHuman2025}. Ither options involve specialized software \cite{damsgaardAnalysisMusculoskeletalSystems2006}, which can not be run on wearables. Therefore, most of modern signal translation solutions rely on machine learning.

There are several works concerning time-series translation, but only a few directly deal with inertial signals. In \cite{wuAutoKeyUsingAutoencoder2020}, the autoencoders are used to predict the output of a paired sensor worn in another body area to generate keys protecting the body area network from eavesdropping. The autoencoders are also utilized in \cite{changUnifiedAutoencoderMethod2023} to tackle person recognition problems originating from inconsistencies between signals registered using sensors worn differently.

Inertial signal translation for motion analysis was presented in \cite{soonUsingSmartwatchData2023}, where smartphone-based measurement results were translated to ones collected in the lower-body area using neural networks. The models used for that purpose included feed-forward dense neural networks and convolution-Long Short-Term Memory (CNN-LSTM) architectures. Although the translation was very accurate, the proposed models were large (more than 12 million parameters), and it might be problematic to implement them on constrained devices.

Given the rising popularity of generative methods, recent works employ Generative Adversarial Networks (GANs) for signal translation and generation. In \cite{wangScaleDirectionGuided2024}, GANs are used to transform noisy inertial signals recorded with a cheaper sensor into those that would be obtained using a more expensive one. Another example of GAN-based inertial signal generation is presented in \cite{CGANbasedHighDimensional2025}, where GANs are used to generate inertial signals corresponding to selected physical activities.

The work presented in the following paper is the continuation of research described in \cite{kolakowskiWristtoTibiaShoeInertial2024}, where the inertial signals were translated between the ones obtained with wrist-worn devices and devices strapped to shins and shoes. The previously tested neural network architectures included a dense neural network, convolutional autonecoder (CNN AE), CNN-based U-Net, and an LSTM network trained in a supervised setting. The main contributions of the paper are the following:
\begin{itemize}
	\item The Conditional Generative Adversarial Networks architectures are tested in an inertial signal translation scenario.
	\item In addition to translating all signals, a different scenario, in which only two required components (mediolateral component of angular speed and its total value), is tested.
\end{itemize}

The rest of the paper is organized as follows. Section \ref{sec:concept} describes the inertial signal translation scenario. The employed ML models and experiment results are presented in Sections \ref{sec:models} and \ref{sec:experiments}, respectively. Section \ref{sec:conclusions} concludes the paper.

\section{Gait analysis using inertial signals}
\label{sec:concept}
\begin{figure}[tbp]
	\centerline{\includegraphics[width=\linewidth]{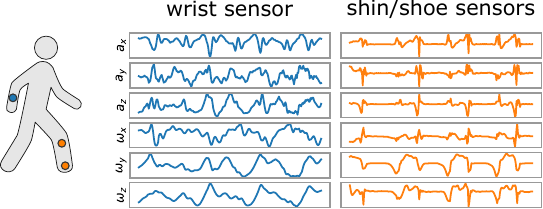}}
	\caption{A difference between inertial signals registered with wrist and shin/shoe sensors.}
	\label{fig:concept}
\end{figure}

In a typical scenario, a gait sensor is equipped with an IMU containing a 3-axis accelerometer and 3-axis gyroscope, resulting in six signal components that can be used for gait analysis.	 Exemplary signals recorded using such sensors are presented in Fig. \ref{fig:concept}.

There is a significant difference between the signals recorded using wrist sensors and sensors placed in the lower body areas. Due to large forces during heel strikes, the acceleration and angular velocity changes are much more abrupt. It enables more precise timing measurements than in the case of wrist signals.
In the proposed concept, the signals registered using a wrist-worn sensor are translated to signals that would be obtained if the sensor was worn in the lower-body area. It enables the use established gait analysis methods without specialized shoe/shin gait sensors. 

In the paper, two approaches are tested. In the first one, all of the signal components are translated, yielding a complete set of  IMU measurement results. The second approach assumes generating only two signals: the angular velocity in the mediolateral axis (ankle rotation axis, where, according to gait studies \cite{tuncaInertialSensorBasedRobust2017}, the most prominent movements happen - $\omega_y$ in Fig. \ref{fig:concept}) and the total value of the angular velocity $\omega_{\mathrm{tot}}= \sqrt{\omega_x^2+ \omega_y^2+\omega_z^2}$. In most cases, those two components are sufficient for detecting heel strike, flat foot, and toe-off phases of the gait cycle \cite{tuncaInertialSensorBasedRobust2017}.

\section{Conditional GAN-based signal translation}
\label{sec:models}
\subsection{Training set-up}

\begin{figure}[tbp]
	\centerline{\includegraphics[width=\linewidth]{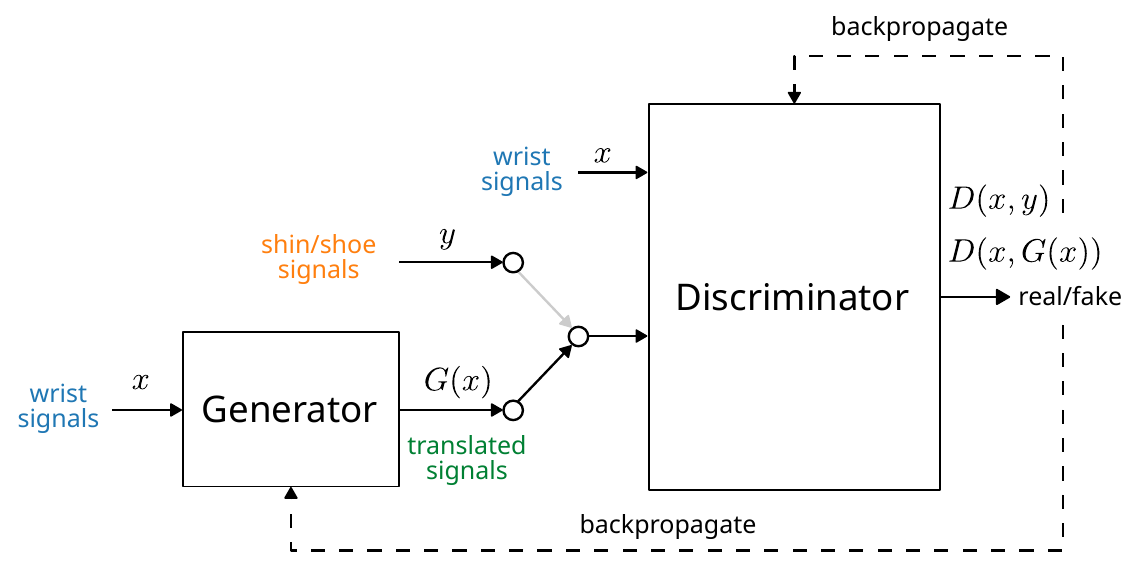}}
	\caption{Training a conditional GAN network to translate signals between sensor locations.}
	\label{fig:gan_training}
\end{figure}

The signal translation is performed using conditional GAN networks. A typical GAN network consists of two separate models, the Generator and the Discriminator, trained in an adversarial manner as presented in Fig. \ref{fig:gan_training}.

The architecture proposed for the IMU signal translation task follows the guidelines outlined in the seminal conditional GAN paper \cite{mirzaConditionalGenerativeAdversarial2014}. The Generator network is responsible for generating (translating) new data based on supplied labels (in this case, a set of six inertial signals). The Discriminator network then evaluates both real and translated signals, judging whether they they are real or fake. To evaluate translation quality, the Discriminator also takes in the original set of signals. The classification errors are backpropagated to update the weights of both the Discriminator and the Generator.

In classical setup, the networks are trained in an adversarial manner, trying to minimize (with respect to the Discriminator) and maximize (with respect to the Generator) the adversarial loss function $L_{\mathrm{adv}}$:	
\begin{multline}
	L_{\mathrm{adv}} = \mathbb{E}_{y\sim p_{\mathrm{data}}}\log\left[D(x, y)\right] \\+ \mathbb{E}_{y\sim\mathrm{model}}\log \left[1-D(x, G(x))\right]
\end{multline}
where $\mathbb{E}_{y\sim p_{\mathrm{data}}}$ and $\mathbb{E}_{y\sim\mathrm{model}}$ are expected values with respect to the real and generated samples respectively; $G(x)$, $D(x,y)$  are Generator and Discriminator outputs; $x$ are inertial signals registered using the wrist sensor and $y$ is a corresponding label (shin/shoe signals).

The paper also tests another approach - Wasserstein GANs with gradient penalty (GP-WGANs). In such a setup, the networks have different loss functions $L_{\mathrm{G}}$ and $L_{\mathrm{D}}$:	
\begin{gather}
	L_{\mathrm{G}}  = - \frac{1}{N}\sum_i D(x_i, G(x_i))\\
	L_{\mathrm{D}} = \frac{1}{N}\sum_i D(x_i, G(x_i)) - \frac{1}{N} \sum_i D(x_i, y_i) + \lambda L_{\mathrm{GP}}\\
	L_{\mathrm{GP}} = \frac{1}{N} \sum_i ((\|\nabla_{\breve{y}_i}D(x_i, \breve{y}_i)\|_2 - 1)^2) \\
	\breve{y}_i = \alpha y_i + (1-\alpha)G(x_i) \quad \textrm{where} \quad \alpha\sim\mathcal{U}[0,1]
\end{gather}
The Generator loss ($L_{\mathrm{G}}$) depends on how well the discriminator performs when classifying generated signals $G(x_i)$.
The loss for the Discriminator ($L_{\mathrm{D}}$) includes an additional gradient penalty computed based on its performance on interpolated signals $\breve{y}_i$ being a randomly weighted sum of the real and generated examples. The influence of the gradient penalty can be controlled using the $\lambda$ parameter.

	\subsection{Proposed generator and discriminator architectures}
	Gait analysis requires a relatively high sampling rate (at least 50 Hz but typically 100 Hz or more) resulting in large volumes of data. As transferring large volumes of data is impractical, in real-use scenarios, it would be preferable to translate the signals and perform the analysis locally on the wrist sensor. Therefore, the translation models should have a limited complexity to optimize the processing time and energy efficiency. As most available AI-enabled chips are optimized for CNNs, the tested generator architectures are CNN-based without using the recurrent layers typically employed for temporal data processing. The details of the generator and discriminator architectures are presented in Tables \ref{tab:gen_architectures} and \ref{tab:disc_arch}.

			\begin{table}[tbp]
		\caption{Generator networks architectures}
		\begin{center}
			\renewcommand{\arraystretch}{1.2}
			\begin{tabular}{ll|ll}
				\hline
				\multicolumn{2}{c|}{\textbf{Autoencoder Generator}}& \multicolumn{2}{|c}{\textbf{U-Net Generator}}\\
				\hline
				\textbf{Layer}&\textbf{Output size} & \textbf{Layer}&\textbf{Output size}\\
								\hline
Input	&	(1, 6, 256) 	&	Input	&	(1, 6, 256) \\
Conv. 1D	&	(1, 64, 256) 	&	Conv. 1D	&	(1, 64, 256) \\
LeakyReLU	&	(1, 64, 256) 	&	LeakyReLU: x1	&	(1, 64, 256) \\
MaxPool1D	&	(1, 64, 128) 	&	MaxPool1D	&	(1, 64, 128) \\
Conv. 1D	&	(1, 128, 128)  	&	Conv. 1D	&	(1, 128, 128)  \\
LeakyReLU	&	(1, 128, 128)  	&	LeakyReLU: x2	&	(1, 128, 128)  \\
MaxPool1D	&	(1, 128, 64)  	&	MaxPool1D	&	(1, 128, 64)  \\
Conv. 1D	&	(1, 256, 64)  	&	Conv. 1D	&	(1, 256, 64)  \\
LeakyReLU	&	(1, 256, 64)  	&	LeakyReLU	&	(1, 256, 64)  \\
MaxPool1D	&	(1, 256, 32)  	&	ConvTrans. 1D	&	(1, 128, 128)  \\
ConvTrans.1D	&	(1, 256, 64)  	&	LeakyReLU	&	(1, 128, 128)  \\
LeakyReLU	&	(1, 256, 64)  	&	Concat(x, x2)	&	(1, 256, 128)  \\
ConvTrans. 1D	&	(1, 256, 128)  	&	ConvTrans. 1D	&	(1, 64, 256)  \\
LeakyReLU	&	(1, 256, 128)  	&	LeakyReLU	&	(1, 64, 256)  \\
ConvTrans.1D$^{\mathrm{a}}$	&	(1, N, 256)  	&	Concat(x, x1)	&	(1, 128, 256)  \\
Sigmoid$^{\mathrm{a}}$	&	(1, N, 256)  	&	Conv. 1D	&	(1, N, 256)  \\
&		&	Tanh	&	(1, N, 256)  \\
\hline
Parameters no.	&	247 938	&	Parameters no.	&	272 898\\
				\hline
				\multicolumn{4}{l}{$^{\mathrm{a}}$ N is the number of translated signal channels (6 or 2)}
			\end{tabular}
			\label{tab:gen_architectures}
		\end{center}
	\end{table}
	
For the generator, two architectures were tested: Convolutional Autoencoder (CNN AE) and U-Net. The generator models, as an input, accept six channel time series comprising acceleration and angular velocity measured with the wrist sensor. The input signals are scaled into the 0--1 range. The input sequence length is 256 samples corresponding to 5.12 s (the sampling rate was set to 50 Hz). The network contains solely convolutional layers followed with Leaky ReLU activation (negative slope parameter equal to 0.2) and MaxPooling. Depending on the translation scenario, the generators output six or two channels. The size of the models is moderate and does not exceed 300,000 parameters.

The Discriminator is a convolutional network following the pattern similar to the generators. To obtain better generalization, a 20 percent dropout is used.

	\section{Experimental verification}
\label{sec:experiments}
The proposed setup was experimentally tested based on  data gathered with custom sensors developed at Warsaw University of Technology (WUT).  The devices were equipped with Bosch Sensortec BMI270 IMUs allowing for 3-axial acceleration (in $\pm 4 \mathrm{g}$ range) and angular velocity (in $\pm 2000 \mathrm{dps}$ range) measurements. The synchronization between the signals was attained by triggering the data collection start over the BLE-radio link (the devices were controlled with BLE-enabled nRF52833 Nordic Semiconductor microcontrollers).

The data used in the study is publicly available at \cite{kolakowskiWristTibiaShoe2023}. The dataset contains 856 examples comprised of wrist-shoe inertial signal sequence pairs of 5.12 s length (50 Hz sampling rate). Both input and output signals were scaled into the 0--1 range. The data were divided into training and validation datasets following a 0.9--0.1 split. The proposed network architectures were then trained using the the parameters listed in Table \ref{tab:disc_arch}.

\begin{figure*}[htbp]
	\centerline{\includegraphics[width=0.97\linewidth]{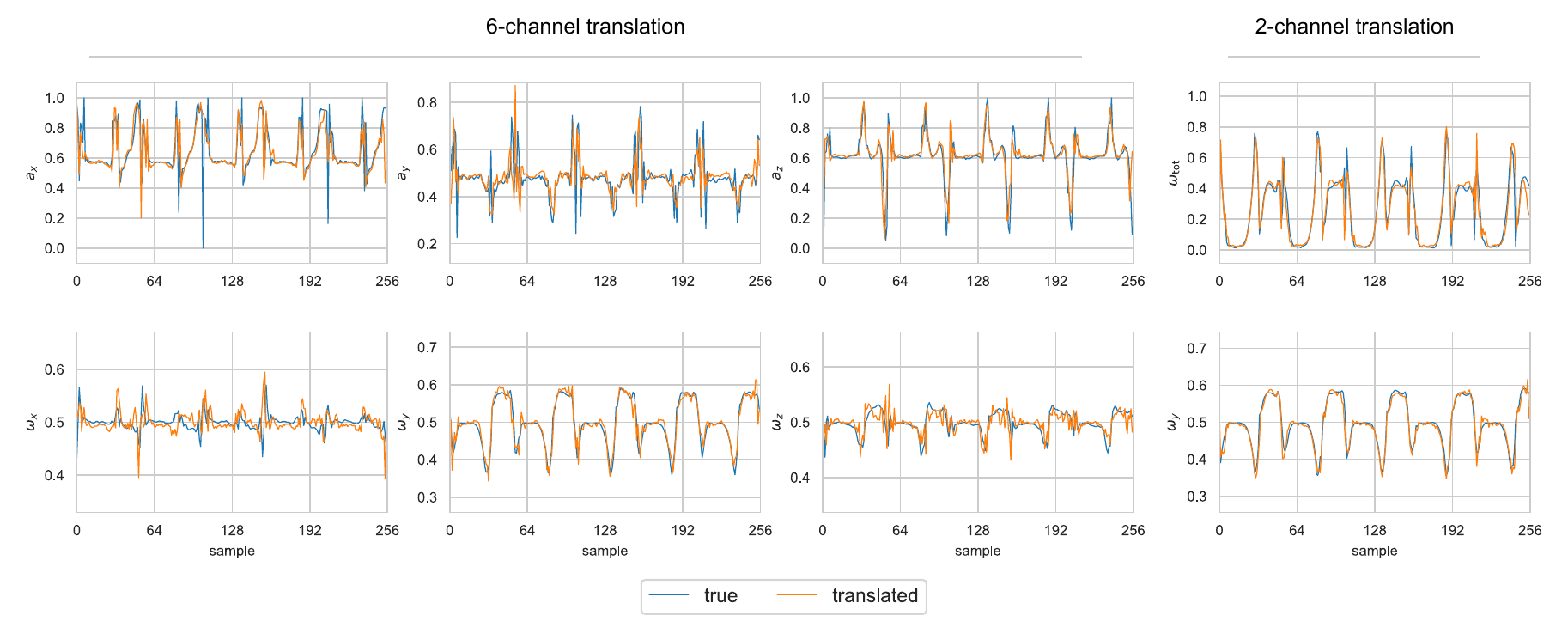}}
	\caption{Exemplary translation results for the 6-channel autoencoder WGAN and the 2-channel U-Net GAN.}
	\label{fig:exemplary_results}
\end{figure*}

The translation accuracy was evaluated based on the root mean squared error (RMSE) and mean absolute error (MAE). The obtained values are listed in Table \ref{tab:metrics}. The selected translated signals are presented in Fig. \ref{fig:exemplary_results}.

			\begin{table}[tbp]
	\caption{The Discriminator Network architecture and training parameters}
	\begin{center}
		\renewcommand{\arraystretch}{1.2}
		\begin{tabular}{ll|ll}
			\hline
			\multicolumn{2}{c|}{\textbf{Discriminator}}& \multicolumn{2}{|c}{\textbf{Training parameters}}\\
			\hline
			\textbf{Layer}&\textbf{Output size} & \textbf{Parameter}&\textbf{Value}\\
			\hline
	Input$^{\mathrm{a}}$	&	(1,6+N,256)	&	epochs	&	10,000\\
Conv. 1D	&	(1, 64, 254) 	&	optimizer	&	Adam\\
LeakyReLU	&	(1, 64, 254) 	&	learning rate GAN	&	$1\mathrm{e}{-4}$\\
MaxPool1D	&	(1, 64, 127) 	&	learning rate WGAN	&	$3\mathrm{e}{-5}$\\
Conv. 1D	&	(1, 64, 125) 	&	Gradient Penalty $\lambda$	&	15\\
LeakyReLU	&	(1, 64, 125) 	&		&	\\
MaxPool1D	&	(1, 64, 62) 	&		&	\\
Conv. 1D	&	(1, 64, 60) 	&		&	\\
LeakyReLU	&	(1, 64, 60) 	&		&	\\
MaxPool1D	&	(1, 64, 30) 	&		&	\\
Flatten	&	(1, 7680)	&		&	\\
Linear	&	(1, 128)	&		&	\\
LeakyRelu	&	(1, 128)	&		&	\\
Dropout	&	(1, 128)	&		&	\\
Linear	&	(1, 1)	&		&	\\
Sigmoid	&	(1, 1)	&		&	\\
			\hline
			Parameters no.	&	247 938	&	&\\
			\hline
							\multicolumn{4}{l}{$^{\mathrm{a}}$ N is the number of translated signal channels (6 or 2)}
		\end{tabular}
		\label{tab:disc_arch}
	\end{center}
\end{table}

	 \begin{table*}[tbp]
		\caption{Wrist to shoe-sensor signal translation errors}
		\begin{center}
			\renewcommand{\arraystretch}{1.2}
			\addtolength{\tabcolsep}{-0.2em}
			\begin{tabular}{|l|cccccc|cccccc|cc|cc|}
				\hline
				\multirow{ 3}{*}{\textbf{model}} &\multicolumn{12}{c|}{\textbf{6-channel }}&\multicolumn{4}{c|}{\textbf{2-channel}}\\
				\cline{2-17}
				&\multicolumn{6}{c|}{\textbf{RMSE}}&\multicolumn{6}{c|}{\textbf{MAE}}&\multicolumn{2}{c|}{\textbf{RMSE}}&\multicolumn{2}{c|}{\textbf{MAE}}\\
				\cline{2-17}
				&$a_x$&$a_y$&$a_z$&$\omega_x$&$\omega_y$&$\omega_z$&
				$a_x$&$a_y$&$a_z$&$\omega_x$&$\omega_y$&$\omega_z$ &
				$\omega_{\mathrm{tot}}$&
				$\omega_y$&$\omega_{\mathrm{tot}}$ &$\omega_y$\\
				\hline
				CNN AE GAN & 0.133 & 0.080 & 0.102 & 0.022 & 0.030 & 0.025 & 0.069 & 0.047 & 0.055 & 0.015 & 0.021 & 0.018 & 0.099 & 0.023 & 0.062 & 0.015 \\
				CNN AE WGAN & 0.123 & 0.078 & 0.099 & 0.024 & 0.029 & 0.018 & 0.065 & 0.045 & 0.053 & 0.016 & 0.020 & 0.012 & 0.182 & 0.090 & 0.129 & 0.071 \\
				U-Net GAN & 0.123 & 0.082 & 0.095 & 0.020 & 0.027 & 0.019 & 0.062 & 0.045 & 0.050 & 0.013 & 0.018 & 0.012 & 0.096 & 0.024 & 0.061 & 0.015 \\
				U-Net WGAN& 0.142 & 0.087 & 0.114 & 0.049 & 0.047 & 0.032 & 0.078 & 0.051 & 0.067 & 0.039 & 0.032 & 0.020 & 0.116 & 0.040 & 0.075 & 0.029 \\
				\hline
				\multicolumn{17}{l}{$^{\mathrm{a}}$The error values are presented for the output 0--1 range. Acceleration and angular velocity were measured in $\pm 4 \mathrm{g}$  and  $\pm 2000 \mathrm{dps}$ ranges. }\\

			\end{tabular}
			\label{tab:metrics}
		\end{center}
	\end{table*}

	For the 6-channel translation, the best results were achieved using the CNN AE WGAN network closely followed by the U-Net GAN. The translation errors were larger for the acceleration signals due to higher noise present in such measurements. In the case of the 2-channel translation, the mediolateral angular velocity component ($\omega_y$) reconstruction accuracy was slightly better than when translating all of the signals. 
	
	The visual analysis of the translated signals shows that, as in the earlier study \cite{kolakowskiWristTibiaShoe2023}, the networks struggle with reconstruction of sharp signal peaks (especially noticeable for the acceleration signals). However, the GAN-based reconstruction of crucial signal areas (like valleys in $\omega_y$ indicating toe-offs and heel-strikes) is much better which would lead to more accurate gait timing analysis.
	
\section{Conclusions}
			\label{sec:conclusions}
In the paper, conditional GAN networks were tested in the scenario of inertial signal translation between two sensor locations. The tested networks comprised convolutional networks of moderate complexities. The experiment results have shown that the translation accuracy is sufficient for the signals to be used in gait analysis scenarios. Using the GAN training setup allows the translating network to better grasp the shoe sensor signal characteristics which leads to better results than in the case of standard autoencoder networks \cite{kolakowskiWristTibiaShoe2023}.

The translation errors  might be reduced by enlarging the dataset or using semi-supervised methods. Another possible development direction is to introduce loss functions taking into account the timing between the  moments corresponding to typical gait events in both real and translated signals.


\end{document}